\documentclass[reprint,amsmath,amssymb,aps,prc,nofootinbib]{revtex4-2}
\usepackage{bm}
\usepackage{graphicx}
\usepackage{dcolumn}
\usepackage{color}
\newcommand{\be}{\begin{equation}}
\newcommand{\ee}{\end{equation}}
\newcommand{\bea}{\begin{eqnarray}}
\newcommand{\eea}{\end{eqnarray}}

\begin{document}

\title{ Description of the odd $^{249-253}$No nuclei using Skyrme functionals
with modified spin-spin interaction.
}
\date{\today}
\author{N. Lyutorovich}
\affiliation{St. Petersburg State University, St. Petersburg, 199034,
  Russia}
\email{n\_lyutor@mail.ru}

\date{\today}

\begin{abstract}
The Skyrme energy-density functionals with modified spin-spin interaction
and pairing strength are used for description of the $^{249}$No, $^{251}$No,
and $^{253}$No excited states.
The results of the HFB and cranked HFB calculations taking into account
the blocking effect are in reasonable agreement with the available data.
For many states, including rotational and three-quasiparticle states,
the results of self-consistent calculations were obtained for the first time.
\end{abstract}

\maketitle
%
\section{Introduction}
%
Self-consistent methods based on the energy-density functional (EDF)
theory are very important and, to some extent, are the only ones
available for the microscopic description of both various nuclei,
including nuclei far from the stability valley, and the properties
of nuclear matter in astrophysical problems. The Skyrme EDFs used
in the nonrelativistic self-consistent approaches contain terms that
are bilinear both in time-even (T-even) and in time-odd (T-odd)
densities and currents: see Refs.~\cite{Bender_RMP_2003,Engel_1975}.
The T-odd densities and currents are equal to zero in the ground states
of the even-even nuclei; therefore, the T-odd EDF terms do not affect
the ground-state properties of these nuclei. The words "time-odd" mean,
as usual, that the EDF terms are constructed from odd densities and
currents, while the EDFs themselves are T-even.
The T-odd EDF terms impact on the characteristics of the excited states
of the even-even nuclei and on the ground- and excited-state properties
of the odd- and odd-odd nuclei.

Recently, the Skyrme functionals SV-bas$_{-0.44}$ and UNEDF1 containing
T-odd spin terms were obtained in Refs.~\cite{Tselyaev_2020} and
\cite{Sassarini_2022}, respectively. The functional SV-bas$_{-0.44}$
obtained on the base of the SV-bas parameter set~\cite{Kluepfel_2009}
had the spin-spin terms constructed from the Landau-Migdal interaction
with the parameters $g$ and $g'$. The $g,g'$ and new spin-orbit
parameters were fitted to reproduce the basic experimental characteristics
of the M1 excitations in $^{208}$Pb within the renormalized time-blocking
approximation and, at the same time, to describe the nuclear ground-state
properties with approximately the same accuracy as the original SV-bas set.

The modified functional UNEDF1 was obtained in Ref.~\cite{Sassarini_2022}
(see, also, \cite{Bonnard_2023} and references therein).
The EDF differs from the original one~\cite{Kortelainen_2012}
by the spin-spin interaction implemented in terms of the Landau-Migdal
parameters $g$ and $g'$ that were adjusted to describe the magnetic
moments of the odd nuclei. Later, in Ref.~\cite{Gray_2023},
it was assigned as "UNEDF1 with T-odd".

The predictive power of the SV-bas$_{-0.44}$ was tested
in the description of excited states in $^{40}$Ca, $^{90}$Zr,
and $^{208}$Pb excited states: see Ref.~\cite{Lyutorovich_2023}.
The modified UNEDF1 functional was employed and tested in calculations
of the electromagnetic moments in states of the odd neighbors
of doubly magic nuclei and of the deformed odd nuclei with
$63 \le Z\le 82$ and $82 \le N \le 126$:
see Refs.~\cite{Sassarini_2022,Bonnard_2023,Gray_2023}.
At the same time, both the modified functionals were not used
for superheavy nuclei.

The purpose of this article is to describe the excited states
of odd $^{249-253}$No nuclei within the framework
of the Skyrme-Hartree-Fock-Bogolyubov (SHFB) and
cranking SHFB approaches and to employ the modified UNEDF1
and SV-bas$_{-0.44}$ functionals in the calculations.
We choose $^{249,251,253}$No because new experimental
data have been obtained for these nuclei in recent years.
In addition, $^{253}$No is one of the heaviest nuclei for which
rather detailed experimental information is available.

There are numerous theoretical studies for $^{249,251,253}$No, but most
of them are based on phenomenological versions of the mean field
and the residual interaction and are thus not self-consistent.
In particular, the microscopic-macroscopic model with the Woods-Saxon
potential and the monopole pairing were used
in Refs.~\cite{Cwiok_1994,Parkhomenko_2005}.
Calculations in the framework of the quasiparticle-phonon model (QPM)
were performed in Refs.~\cite{Adamian_2011,Shirikova_2015,Adamian_2018}
and, in addition, in Ref.~\cite{Adamian_2011} results were obtained
within the framework of the microscopic-macroscopic two-center shell model.

There are few self-consistent calculations for these nuclei, but their
results are not entirely consistent with each other.
The SHFB method combined with Lipkin-Nogami approximation (SHFB+LN)
were used in Refs.~\cite{Bender_NPA_2003,Adamian_2011}.
Calculations in the framework of SHFB for several EDFs and
the relativistic Hartree-Bogolyubov calculations were performed
in Refs.~\cite{Dobaczewski_2015}. The relativistic mean-field approach
was also used in Ref.~\cite{Singh_2021} to investigate ground-state
properties of many transfermium nuclei,including nobelium isotopes.

Comparing the results of the self-consistent calculations, it can be noted
that the energies ($E$) of the $^{253}$No levels calculated with the SLy4
Skyrme interaction in Ref.~\cite{Adamian_2011,Adamian_2021}
are by 600--700 keV less than similar values in Ref.~\cite{Dobaczewski_2015}
and approximately two times less the values in Ref.~\cite{Bender_NPA_2003}.
The difference between values \cite{Dobaczewski_2015} and \cite{Bender_NPA_2003}
is not very significant and may be explained by the approximations used.
At the same time, the low $E$ values~\cite{Adamian_2011,Adamian_2021}
may be explained by incorrect pairing strength in the SHFB calculations.
In this work, the SHFB+LN approximation was used for the SLy4 and SkP
parameter sets but the original (for these sets) pairing strength
was adjusted without the LN method.
The fact is that the pairing parameters for the SHFB and SHFB+LN are
significantly different: see, for example, Refs.~\cite{Shi_2014}, p. 3
and \cite{Bonnard_2023}, p. 3.
Using the latest version \cite{Dobaczewski_2021} of the code HFODD used
in Ref.~\cite{Adamian_2011}, one can check that the $E$ values
Refs.~\cite{Adamian_2011,Adamian_2021} for SLy4 and SkP parameter sets
are not correct while these EDFs with correct pairing strengths
give results that are in agreement with
Refs~\cite{Dobaczewski_2015} and \cite{Bender_NPA_2003}. 
This, of course, does not reduce the value of other
results of Refs.~\cite{Adamian_2011,Adamian_2021} that were obtained
in the framework of the models based on the phenomenological mean
and pairing fields, i.e., the QPM approach and  mic-mac two-center shell model.

Thus, to date, most calculations for $^{249,251,253}$No were based
on the phenomenological mean fields. There are only few self-consistent
calculations for the excited states in $^{253}$No however their results
differ significantly from experimental data. The given paper presents
the investigation of the $^{249}$No, $^{251}$No, and $^{253}$No excited
states performed in the framework of the SHFB and cranked SHFB methods
with implementation of the modified SV-bas$_{-0.44}$ and UNEDF1 Skyrme
functionals. Calculations were performed using a slightly modified
version of the code HFODD~\cite{Dobaczewski_2021}.

The paper is organized as follows. Calculation details are given in
Section~\ref{sec:theor}. The neutron and proton pairing strengths for
the modified functionals are adjusted to reproduce the experimental
odd-even mass differences for actinide nuclei in Subsec.~\ref{pairing}.
The results of calculations for $^{253}$No, $^{251}$No, and $^{249}$No
are presented in Subsections~\ref{253No}, \ref{251No}, and \ref{249No}.
Conclusions are given in Sec.~\ref{sec:conclusions}.

To distinguish the modified SV-bas$_{-0.44}$ and UNEDF1 functionals 
from the original ones, we will denote them as bas44 and UDF1m,
keeping in mind that in addition to the T-odd terms they have also
new pairing parameters.
%
\section{Calculation details} \label{sec:theor}
The Skyrme EDF theory was described in many publications:
see, for example, the reviews~\cite{Bender_RMP_2003,Schunck_2019}.
In the given paper, the equations of the theory are solved using
the version 3.18j (see Ref.~\cite{Bonnard_2023}) of the code HFODD
with slight modifications. The code has been developed over many years
and has been described in a number of publications, which also provide
detailed formulas: see Ref.~\cite{Dobaczewski_2021} and references therein.

It is known (see, for example, Ref.~\cite{Kortelainen_2012}, p. 10),
that the single-particle levels close to the Fermi surface affect 
the pairing correlations, therefore, when changing EDFs, it is necessary
to modify the pairing terms. The pairing interaction was not used
for the SV-bas$_{-0.44}$ and modified UNEDF1~\cite{Sassarini_2022} functionals.
In Ref.~\cite{Bonnard_2023}, the pairing strength of the modified UNEDF1
functional was determined approximately: the neutron and proton pairing
parameters were increased by 20\% with respect to the original EDF
to compensate for the effects of the LN method that was not used
in the calculations ~\cite{Bonnard_2023} (the original EDF UNEDF1
was fitted in the framework of the LN method). Since correct determination
of pairing forces is very important for a self-consistent description
of excited states in deformed nuclei, new pairing forces are obtained
below for both the modified functionals.

The pairing terms in the EDFs have the form
\be \label{eq:E_pair}
E^{pair} = \int d^3 {\mathbf r}\, \tilde\chi ({\mathbf r})\,.
\ee
The pairing energy density $\tilde\chi ({\mathbf r})$ is derived
from the mixed pairing force in which the volume and surface components of
density-dependent delta interaction are mixed in the equal proportions
(see Ref.~\cite{Kortelainen_2012}):
\be \label{eq:chi}
 \tilde\chi ({\mathbf r})
 = \sum_{q=n,p} \, \frac{V_q}{2} \left[1-\frac{\rho_0(\mathbf r)}{2\rho_c}\right]
 \tilde\rho^2_q ({\mathbf r})\, ,
\ee
where $\tilde\rho_c$ is the local pairing density, $\tilde\rho_c = 0.16$ fm$^{-3}$.
The neutron and proton pairing strengths were adjusted to match
the experimental odd-even mass differences, for which the three-point parameters
$\Delta^{(3)}_n$ and $\Delta^{(3)}_p$ centered at odd particle numbers were used.
The equation for the neutron $\Delta^{(3)}_n$ value reads
\be \label{eq:D3}
\Delta^{(3)}_n (N) = 1/2 [B(Z,N-1)+B(Z,N+1)-2B(Z,N)],
\ee
where $B(Z, N)$ value is the binding energy of the nucleus with the even
number of protons $Z$ and odd number of neutrons $N$.
The equation for the proton value $\Delta^{(3)}_p (N)$ has the similar form.

The rotational symmetry is taken into account approximately, within
the framework of the cranking HFB method: see Ref.~\cite{Ring_Schuck}.
In this approximation, the EDF contains the term with the constraint
\be \label{eq:J_y}
\langle J_y \rangle = \sqrt{I(I+1)-\langle J_y^2 \rangle},
\ee
where, $I$ is the spin of the nucleus, the $J_y$ and $J_z$ are the operators
of the nuclear angular momentum in the intrinsic frame of reference.
Using the method we calculate the rotational energy and take into the effect
of the rotation on the mean field and pairing.

The methodology of the calculations is similar to one described
in Ref.~\cite{Bonnard_2023} but with some modifications.
The SHFB equations were solved by expanding the mean-field wave functions
on spherical harmonic-oscillator (HO) basis states including $N=16$
quanta in either of the Cartesian directions that gave 969 states.
To study any possible effect of the HO space on the convergence
of the iteration procedure and on the final results, calculations
were also performed using the deformed HO basis of 959 states and
the HO deformation close to the final deformation of the nucleus ground state.
The HFB calculations were performed with the cutoff of $E_{\mbox{cut}} = 60$ MeV
used to truncate the quasiparticle space.

First, the SHFB calculations for the given nucleus were performed
constraining the axial mass quadrupole moment to $Q_{20}=34$ b and,
then, the calculations were performed without the constraint.
In the second step, the HFB properties of the nucleus states, including
total energies, deformations, pairing characteristics and others,
were obtained by blocking the relevant quasiparticle levels.
The diabatic blocking and the Broyden method were used to ensure
the convergence of the iteration procedure. 
In the third step, the total HFB states obtained in the second stage
were used to take into account the nuclear rotation in the framework
of the cranking HFB approximation. The excitation energies were calculated
as differences of the total energies of the states with different
blocked quasiparticle states.
%
\section{Results} \label{sec:results}
%
\subsection{Pairing strength} \label{pairing}
%
The pairing force parameters $V_n$ and $V_p$ were determined by fitting
the calculated 3-point odd-even mass differences $\Delta^{(3)}$,
defined by Eq.~\ref{eq:D3}, to experimental values for nuclei
of the actinides region. In the fitting, the theoretical binding energies
$B(Z, N)$ were calculated in the HFB approximation taking into account
the blocking effect and the experimental $B(Z,N)$ values were taken from Ref.~\cite{nudat3}.
The $\Delta^{(3)}_p$ values centered at odd-proton numbers were used
for the $^{237}$Np, $^{243}$Am, and $^{249}$Bk nuclei.
The $V_p$ parameter was fitted for each of these nuclei and, after that,
a weighted mean value was obtained taking into account experimental uncertainties.
The $\Delta^{(3)}_n$ values centered at odd-neutron numbers
were used for the $^{233}$Th, $^{251}$Cf, and $^{253}$No, but the $^{253}$No
value has too much experimental uncertainty so it's input to the weighted
mean $V_n$ value was very small.
As a result of the fittings, the following pairing parameters were obtained:\\
$(V_n,V_p) = (-228.4,\,-270)$ MeV, for UDF1m,\\
$(V_n,V_p) = (-252.9,\,-306)$ MeV, for bas44.
%
\subsection{$^{253}$No} \label{253No}
%
All the results of the calculations for $^{253}$No using the UDF1m
functional and available experimental data are presented
in Table~\ref{table:253No_levels}.
Here, $I^\pi$, and $E$  denote the level spin, parity, and energy (in keV),
respectively. The configurations of the nucleus states are determined
by the structure of blocked one- and three-quasiparticle (1qp and 3qp) states
for which the dominant component in the wave function is labeled
by the asymptotic quantum numbers $\Omega[Nn_z\Lambda]$ of the blocked qp state.
The $K$ value is a projection of the nucleus spin on the symmetry axis:
$K=\Omega$ for a 1qp state and $K=\sum_i \Omega_i$ for a 3qp state.
The experimental data were taken from Refs.~\cite{A=251-259,ENSDF,Lopez-Martens_2007}.
The calculations show that the UDF1m and bas44 functionals give rather similar
results but the results for UDF1m are in some better agreement
with the data therefore the table presents only the UDF1m results.
A comparison of the EDFs will be given below, in the table for $^{251}$No.
\begin{table}[t]
\caption{\label{table:253No_levels}
$^{253}$No states obtained in the HFB and CHFB calculations
for the Skyrme EDF UDF1m in comparison with experimental data.
Here, $I^\pi$ and $E$  denote spin, parity and energy (in keV) of the levels.
Configurations of the blocked 1qp and 3qp states are shown
by dominant components in the wave functions where the asymptotic quantum numbers
$\Omega[Nn_z\Lambda]$ label the blocked qp state.
The $K$ value is a projection of the nucleus spin on the symmetry axis:
$K=\Omega$ for 1qp state and $K=\sum \Omega$ for 3qp state.
The experimental data were taken from Refs.~\cite{A=251-259,ENSDF,Lopez-Martens_2007}
}
\begin{ruledtabular}
\begin{tabular}{llcc}
\multicolumn{2}{c}{Exp.}
                           &    \multicolumn{2}{c}{UDF1m}\\
\cline{1-2}                       \cline{3-4}
\vspace{-2mm}
            &              &                       &       \\
\ \ $ I^\pi$    & \ \ $E$  &   $ I^\pi\; K[Nn_z\Lambda]$
                                                   &  $E$  \\
\hline
\vspace{-2mm}
            &              &                       &       \\
  $(9/2^-)$ &  \ \ \ 0     &   $ 9/2^-\, 9/2[734]$ &   0\ \  \\
 $(11/2^-)$ &   64.0(10)   &   $11/2^-\, 9/2[734]$ &   63  \\
 $(13/2^-)$ &  132.8? (16) &   $13/2^-\, 9/2[734]$ &  138  \\
   $5/2^+$  &  167.5(5)    &   $ 5/2^+\, 5/2[622]$ &  525  \\
 $(15/2^-)$ &  220.0(15)   &   $15/2^-\, 9/2[734]$ &  226  \\
            &  258.2(12)   &   $ 3/2^+\, 3/2[622]$ &  471  \\
 $(17/2^-)$ &  317.5(16)   &   $17/2^-\, 9/2[734]$ &  325  \\
 (7/2+)     &  355 \cite{Lopez-Martens_2007}
                           &   $ 7/2^+\, 7/2[624]$ &  220  \\
 $(19/2^-)$ &  427.7(16)   &   $19/2^-\, 9/2[734]$ &  435  \\
  $(1/2^+)$ &  450.9(12)   &   $ 1/2^+\, 1/2[620]$ &  342  \\
 $(21/2^-)$ &  551.2(17)   &   $21/2^-\, 9/2[734]$ &  557  \\
            &              &   $ 7/2^+\, 7/2[613]$ &  671  \\
  $(1/2^+)$ &  670         &   $ 1/2^+\, 1/2[631]$ &  934  \\
 $(23/2^-)$ &  686.7(18)   &   $23/2^-\, 9/2[734]$ &  692  \\
 $11/2^-$   &  750 \cite{Hauschild_2022}
                           &   $11/2^-\, 11/2[725]$& 1039  \\
 $(25/2^-)$ &  834.2(18)   &   $25/2^-\, 9/2[734]$ &  839  \\
            &              &   $ 7/2^-\, 7/2[743]$ &  847  \\
 $(27/2^-)$ &  994.2(19)   &   $27/2^-\, 9/2[734]$ &  992  \\
 $(29/2^-)$ & 1165.2(19)   &   $29/2^-\, 9/2[734]$ & 1156  \\

\vspace{-2mm}
            &              &                       &       \\
          \multicolumn{4}{c}{3qp states}                    \\
 $(15/2^-)$ &  934.5(15)   &   $15/2^-\; K=15/2$ B1 \footnotemark[1]
                                                   & 1198  \\
 $(17/2^-)$ & 1022.6(16)   &   $17/2^-\; K=15/2$ B1 \footnotemark[1] \footnotemark[2]
                                                   &       \\
            &              &   $ 3/2^-\; K=3/2$ B2 \footnotemark[1]
                                                   & 1205  \\
            &              &   $17/2^-\; K=17/2$ B3 \footnotemark[1]
                                                   & 1357  \\
            &              &   $23/2^+\; K=23/2$ C1 \footnotemark[3]
                                                   & 1341  \\
            &              &   $17/2^+\; K=17/2$ D1 \footnotemark[4]
                                                   & 1347  \\
            &              &   $9/2^+\; K=9/2$ C2 \footnotemark[3]
                                                   & 1411  \\
            &              &   $15/2^+\; K=15/2$ D1 \footnotemark[4]
                                                   & 1437  \\
\end{tabular}
\end{ruledtabular}
\footnotetext[1]{B1, B2, B3: members of the $\pi 7/2[514]
\otimes \pi 1/2[521] \otimes \nu 9/2[734]$ multiplet}
\footnotetext[2]{a member of the rotational band built on the B1 state}
\footnotetext[3]{C1, C2: members of the $\pi 7/2[514]
\otimes \pi 7/2[633] \otimes \nu 9/2[734]$ multiplet}
\footnotetext[4]{D1, D2: members of the $\pi 1/2[521]
\otimes \pi 7/2[633] \otimes \nu 9/2[734]$ multiplet}
\end{table}

The results presented in the table make it possible to draw the following
conclusions about the states of $^{253}$No.
The level with $I^\pi=7/2^+$ was found at the energy $E=355$ keV
in the experimental work~\cite{Lopez-Martens_2007} that agrees, within
uncertainties, with the value $E=379$ keV experimentally obtained for this level
in Ref~\cite{Hessberger_1997}.
Though this level is not mentioned in the latter experimental papers,
the $I^\pi=7/2^+$ state should exist in the energy region corresponding
to values~\cite{Lopez-Martens_2007,Hessberger_1997}:
our calculations predict the $ 7/2^+\, 7/2[624]$ state in $^{253}$No
at $E=220$ and 513 keV for the parameter sets UDF1m and SV-bas$_{-0.44}$,
respectively. The calculations with phenomenological mean fields also
predict this state at low energy: 24~\cite{Cwiok_1994}, 200~\cite{Parkhomenko_2005},
and 50 keV \cite{Adamian_2011,Shirikova_2015,Adamian_2021}.

The presented calculations give the position of the 1/2[631] state
essentially higher than 1/2[620] state: by 590 and 840 keV for UDF1m and ba44 EDFs, respectively.
Similar results were obtained in Refs.~\cite{Adamian_2011,Shirikova_2015,Adamian_2018,Adamian_2021}
using the QPM model with phenomenological mean fields.
Taking all this into account, one can assign the experimental levels
at 450.9 and 670 keV as the states with the dominant 1/2[620] and 1/2[631]
components, respectively.

Experimental data for 3qp states are still scarce: there are only data
for two states of the 3qp band in $^{253}$No (see the table).
The given calculations predict many 3qp states at $E\gtrsim 1$ MeV
that are members of the 3qp multiplets. Only some of them are shown
in Table~\ref{table:253No_levels}. It should be noted that proton pairing
in these 3qp states is weakened compared to 1qp states: for example,
the UDF1m functional gives the average proton gap parameter $\Delta(p)=0.26$ MeV,
so taking into account the particle-number conservation is more important
in such cases than for 1qp states. Using a method that takes into account
the particle-number conservation, e.g. the LN method, will give a negative
contribution to the total 3qp energies and make the excitation energy of the
states slightly lower.
%
\subsection{$^{251}$No} \label{251No}
%
The results of the calculations for $^{251}$No using the UDF1m and
SV-bas$_{-0.44}$ parameter sets and available experimental data~\cite{A=251,ENSDF}
are presented in Table~\ref{table:251No_levels}.
\begin{table}
\caption{\label{table:251No_levels}
The $^{251}$No states obtained in the HFB and CHFB calculations
for the Skyrme EDFs UDF1m and SV-bas$_{-0.44}$ in comparison
with experimental data~\cite{Lopez-Martens_2022,A=251}.
Denotations are the same as in Table~\ref{table:253No_levels}.
}
\begin{ruledtabular}
\begin{tabular}{llccc}
\multicolumn{2}{c}{Exp.}  &  \multicolumn{3}{c}{Theory}  \\
\cline{1-2}                     \cline{3-5}
\vspace{-2mm}
            &             &                     &      &   \\
\ \ $I^\pi$ & \ \ $E$     & $ I^\pi\; K[Nn_z\Lambda]$
                                                & $E$(UDF1m)
                                                       & $E$(bas44) \\
\hline
\vspace{-2mm}
            &             &                     &      &      \\
 $7/2^+$    &  \ \ \ 0    & $ 7/2^+\, 7/2[624]$ &   0\ &   0\ \\
 $(9/2^+)$  &  (60.3(3))  & $ 9/2^+\, 7/2[624]$ &   63 &  58  \\
 $(1/2^+)$  &  106(6)     & $ 1/2^+\, 1/2[631]$ &  566 &  685 \\
 $(9/2^-)$  &  203.6(2)   & $ 9/2^-\, 9/2[734]$ &  140 &  269 \\
            &             & $ 5/2^+\, 5/2[622]$ &  180 &  116 \\
            &             & $ 7/2^-\, 7/2[743]$ &  486 &  299 \\
            &             & $ 1/2^+\, 1/2[620]$ &  685 &  719  \\
            &             & $ 1/2^-\, 1/2[501]$ &  737 &    \\
 $(7/2^+)$  &  917.2(5)?  & $ 7/2^+\, 7/2[613]$ &  980 & 1105 \\
            &             & $ 1/2^-\, 1/2[761]$ & 1169 &    \\
            &             & $11/2^-\, 11/2[725]$& 1399 &    \\
\end{tabular}
\end{ruledtabular}
\end{table}
The UDF1m and bas44 functionals give very close results for the $^{251}$No states.
The theoretical values are in good agreement with the experimental data
excluding the lowest $1/2^+$ state.
In particular, both of the EDFs predict close energies for the $1/2^+\, 1/2[620]$
and $1/2^+\, 1/2[631]$ levels that exceed the experimental value by 0.5 MeV.
However, the assignment of the level 106 keV is given in Refs.~\cite{A=251,ENSDF}
as preliminary and the calculations predict other lowlying states, therefore
a more accurate determination of the spin and parity of the level 206 keV is desirable.

The calculation results remove the uncertainty in identifying
the experimental level of 60.3(3) keV and clearly indicate
that this is a rotational state with the dominant $ 9/2^+\, 7/2[624]$ component.
%
\subsection{$^{249}$No} \label{249No}
%
The results of the HFB calculations for the excited states in $^{249}$No
and available experimental data \cite{Tezekbayeva_2022} are given
in Table~\ref{table:249No_levels} where denotations are the same as
in Table~\ref{table:253No_levels}.
\begin{table}
\caption{\label{table:249No_levels}
$^{249}$No states obtained in the HFB calculations for the Skyrme EDF UDF1m 
in comparison with experimental data.
Denotations are the same as in Table~\ref{table:253No_levels}.
}
\begin{ruledtabular}
\begin{tabular}{llcc}
\multicolumn{2}{c}{Exp.}
                      &  \multicolumn{2}{c}{UDF1m}\\
\cline{1-2}                \cline{3-4}
\vspace{-2mm}
            &         &                     &       \\
\ \ $ I^\pi$& \ $E$   & $ I^\pi\; K[Nn_z\Lambda]$
                                            & $E$   \\
\hline
\vspace{-2mm}
            &         &                     &       \\
  $ 5/2^+$  & \ \ 0   & $ 5/2^+\, 5/2[622]$ &  0\ \ \\
  $ 1/2^+$  & ~125    & $ 1/2^+\, 1/2[631]$ &  254 \\
            &         & $ 7/2^+\, 7/2[624]$ &  143 \\
            &         & $ 7/2^-\, 7/2[743]$ &  229 \\
            &         & $ 9/2^-\, 9/2[734]$ &  424 \\
            &         & $ 1/2^-\, 1/2[501]$ &  509 \\
            &         & $ 5/2^-\, 5/2[503]$ &  929 \\
            &         & $ 3/2^-\, 3/2[501]$ &  982 \\
            &         & $ 1/2^+\, 1/2[620]$ & 1023 \\
\end{tabular}
\end{ruledtabular}
\end{table}
Comparing the results presented in
Tables~\ref{table:253No_levels} -- \ref{table:249No_levels},
one can notice that the energy of the state having a dominant
component $1/2^+\, 1/2[631]$ is significantly less in $^{249}$No
than in $^{251}$No and $^{253}$No. At the same time, the energy
of the $1/2^+\, 1/2[620]$ state is significantly greater
in $^{249}$No than in $^{251}$No and $^{253}$No.
These results make it possible to determine that the experimental
level $1/2^+$, $E \sim 125$ keV, is a state with a dominant component
$1/2^+\, 1/2[631]$, since there are no other theoretical levels
$1 /2^+$ below 1 MeV.
%
\section{Conclusions} \label{sec:conclusions}
%
Fully self-consistent cranked HFB calculations have been performed
for $^{249}$No, $^{251}$No, and $^{253}$No nuclei with modified
versions the Skyrme energy-density functionals UNEDF1 and SV-bas$_{-0.44}$
in which the T-odd terms were constructed from the Landau-Migdal
spin-spin interaction with the parameters $g$ and $g'$. The new neutron
and proton paring strengths were adjusted for both the functionals
to reproduce the experimental odd-even mass differences for a number
of actinide nuclei. To distinguish these functionals from the original
ones, they were designated as UDF1m and bas44.
The calculations performed within the HFB and cranking HFB
approximations for 1qp and 3qp states and rotational bands,
give comparable results for the UDF1m and bas44 functionals.
The results are in reasonable agreement with available experimental data
but the agreement is some better for UDF1m. For many states, including
the rotational and three-quasiparticle states,
the results of self-consistent calculations were obtained for the first time.
\section*{Acknowledgments}
This work was supported by the Russian Foundation for Basic Research,
project number 21-52-12035.  The research was carried out using
computational resources provided by the Computer Center of
St. Petersburg State University. 
\bibliographystyle{apsrev4-2}
\bibliography{nobelium}

\begin{thebibliography}{31}%
\makeatletter
\providecommand \@ifxundefined [1]{%
 \@ifx{#1\undefined}
}%
\providecommand \@ifnum [1]{%
 \ifnum #1\expandafter \@firstoftwo
 \else \expandafter \@secondoftwo
 \fi
}%
\providecommand \@ifx [1]{%
 \ifx #1\expandafter \@firstoftwo
 \else \expandafter \@secondoftwo
 \fi
}%
\providecommand \natexlab [1]{#1}%
\providecommand \enquote  [1]{``#1''}%
\providecommand \bibnamefont  [1]{#1}%
\providecommand \bibfnamefont [1]{#1}%
\providecommand \citenamefont [1]{#1}%
\providecommand \href@noop [0]{\@secondoftwo}%
\providecommand \href [0]{\begingroup \@sanitize@url \@href}%
\providecommand \@href[1]{\@@startlink{#1}\@@href}%
\providecommand \@@href[1]{\endgroup#1\@@endlink}%
\providecommand \@sanitize@url [0]{\catcode `\\12\catcode `\$12\catcode
  `\&12\catcode `\#12\catcode `\^12\catcode `\_12\catcode `\%12\relax}%
\providecommand \@@startlink[1]{}%
\providecommand \@@endlink[0]{}%
\providecommand \url  [0]{\begingroup\@sanitize@url \@url }%
\providecommand \@url [1]{\endgroup\@href {#1}{\urlprefix }}%
\providecommand \urlprefix  [0]{URL }%
\providecommand \Eprint [0]{\href }%
\providecommand \doibase [0]{https://doi.org/}%
\providecommand \selectlanguage [0]{\@gobble}%
\providecommand \bibinfo  [0]{\@secondoftwo}%
\providecommand \bibfield  [0]{\@secondoftwo}%
\providecommand \translation [1]{[#1]}%
\providecommand \BibitemOpen [0]{}%
\providecommand \bibitemStop [0]{}%
\providecommand \bibitemNoStop [0]{.\EOS\space}%
\providecommand \EOS [0]{\spacefactor3000\relax}%
\providecommand \BibitemShut  [1]{\csname bibitem#1\endcsname}%
\let\auto@bib@innerbib\@empty
\bibitem [{\citenamefont {Bender}\ \emph
  {et~al.}(2003{\natexlab{a}})\citenamefont {Bender}, \citenamefont {Heenen},\
  and\ \citenamefont {Reinhard}}]{Bender_RMP_2003}%
  \BibitemOpen
  \bibfield  {author} {\bibinfo {author} {\bibfnamefont {M.}~\bibnamefont
  {Bender}}, \bibinfo {author} {\bibfnamefont {P.-H.}\ \bibnamefont {Heenen}},\
  and\ \bibinfo {author} {\bibfnamefont {P.-G.}\ \bibnamefont {Reinhard}},\
  }\href {https://doi.org/10.1103/RevModPhys.75.121} {\bibfield  {journal}
  {\bibinfo  {journal} {Rev. Mod. Phys.}\ }\textbf {\bibinfo {volume} {75}},\
  \bibinfo {pages} {121} (\bibinfo {year} {2003}{\natexlab{a}})}\BibitemShut
  {NoStop}%
\bibitem [{\citenamefont {Engel}\ \emph {et~al.}(1975)\citenamefont {Engel},
  \citenamefont {Brink}, \citenamefont {Goeke}, \citenamefont {Krieger},\ and\
  \citenamefont {Vautherin}}]{Engel_1975}%
  \BibitemOpen
  \bibfield  {author} {\bibinfo {author} {\bibfnamefont {Y.~M.}\ \bibnamefont
  {Engel}}, \bibinfo {author} {\bibfnamefont {D.~M.}\ \bibnamefont {Brink}},
  \bibinfo {author} {\bibfnamefont {K.}~\bibnamefont {Goeke}}, \bibinfo
  {author} {\bibfnamefont {S.~J.}\ \bibnamefont {Krieger}},\ and\ \bibinfo
  {author} {\bibfnamefont {D.}~\bibnamefont {Vautherin}},\ }\href@noop {}
  {\bibfield  {journal} {\bibinfo  {journal} {Nucl. Phys. A}\ }\textbf
  {\bibinfo {volume} {249}},\ \bibinfo {pages} {215} (\bibinfo {year}
  {1975})}\BibitemShut {NoStop}%
\bibitem [{\citenamefont {Tselyaev}\ \emph {et~al.}(2020)\citenamefont
  {Tselyaev}, \citenamefont {Lyutorovich}, \citenamefont {Speth},\ and\
  \citenamefont {Reinhard}}]{Tselyaev_2020}%
  \BibitemOpen
  \bibfield  {author} {\bibinfo {author} {\bibfnamefont {V.}~\bibnamefont
  {Tselyaev}}, \bibinfo {author} {\bibfnamefont {N.}~\bibnamefont
  {Lyutorovich}}, \bibinfo {author} {\bibfnamefont {J.}~\bibnamefont {Speth}},\
  and\ \bibinfo {author} {\bibfnamefont {P.-G.}\ \bibnamefont {Reinhard}},\
  }\href {https://doi.org/10.1103/PhysRevC.102.064319} {\bibfield  {journal}
  {\bibinfo  {journal} {Phys. Rev. C}\ }\textbf {\bibinfo {volume} {102}},\
  \bibinfo {pages} {064319} (\bibinfo {year} {2020})}\BibitemShut {NoStop}%
\bibitem [{\citenamefont {Sassarini}\ \emph {et~al.}(2022)\citenamefont
  {Sassarini}, \citenamefont {Dobaczewski}, \citenamefont {Bonnard},\ and\
  \citenamefont {Ruiz}}]{Sassarini_2022}%
  \BibitemOpen
  \bibfield  {author} {\bibinfo {author} {\bibfnamefont {P.~L.}\ \bibnamefont
  {Sassarini}}, \bibinfo {author} {\bibfnamefont {J.}~\bibnamefont
  {Dobaczewski}}, \bibinfo {author} {\bibfnamefont {J.}~\bibnamefont
  {Bonnard}},\ and\ \bibinfo {author} {\bibfnamefont {R.~F.~G.}\ \bibnamefont
  {Ruiz}},\ }\href {https://doi.org/10.1088/1361-6471/ac900a} {\bibfield
  {journal} {\bibinfo  {journal} {Journal of Physics G: Nuclear and Particle
  Physics}\ }\textbf {\bibinfo {volume} {49}},\ \bibinfo {pages} {11LT01}
  (\bibinfo {year} {2022})}\BibitemShut {NoStop}%
\bibitem [{\citenamefont {Kl{\"{u}}pfel}\ \emph {et~al.}(2009)\citenamefont
  {Kl{\"{u}}pfel}, \citenamefont {Reinhard}, \citenamefont {B{\"{u}}rvenich},\
  and\ \citenamefont {Maruhn}}]{Kluepfel_2009}%
  \BibitemOpen
  \bibfield  {author} {\bibinfo {author} {\bibfnamefont {P.}~\bibnamefont
  {Kl{\"{u}}pfel}}, \bibinfo {author} {\bibfnamefont {P.~G.}\ \bibnamefont
  {Reinhard}}, \bibinfo {author} {\bibfnamefont {T.~J.}\ \bibnamefont
  {B{\"{u}}rvenich}},\ and\ \bibinfo {author} {\bibfnamefont {J.~A.}\
  \bibnamefont {Maruhn}},\ }\href {https://doi.org/10.1103/PhysRevC.79.034310}
  {\bibfield  {journal} {\bibinfo  {journal} {Phys. Rev. C}\ }\textbf {\bibinfo
  {volume} {79}},\ \bibinfo {pages} {034310} (\bibinfo {year}
  {2009})}\BibitemShut {NoStop}%
\bibitem [{\citenamefont {Bonnard}\ \emph {et~al.}(2023)\citenamefont
  {Bonnard}, \citenamefont {Dobaczewski}, \citenamefont {Danneaux},\ and\
  \citenamefont {Kortelainen}}]{Bonnard_2023}%
  \BibitemOpen
  \bibfield  {author} {\bibinfo {author} {\bibfnamefont {J.}~\bibnamefont
  {Bonnard}}, \bibinfo {author} {\bibfnamefont {J.}~\bibnamefont
  {Dobaczewski}}, \bibinfo {author} {\bibfnamefont {G.}~\bibnamefont
  {Danneaux}},\ and\ \bibinfo {author} {\bibfnamefont {M.}~\bibnamefont
  {Kortelainen}},\ }\href
  {https://doi.org/https://doi.org/10.1016/j.physletb.2023.138014} {\bibfield
  {journal} {\bibinfo  {journal} {Physics Letters B}\ }\textbf {\bibinfo
  {volume} {843}},\ \bibinfo {pages} {138014} (\bibinfo {year}
  {2023})}\BibitemShut {NoStop}%
\bibitem [{\citenamefont {Kortelainen}\ \emph {et~al.}(2012)\citenamefont
  {Kortelainen}, \citenamefont {McDonnell}, \citenamefont {Nazarewicz},
  \citenamefont {Reinhard}, \citenamefont {Sarich}, \citenamefont {Schunck},
  \citenamefont {Stoitsov},\ and\ \citenamefont {Wild}}]{Kortelainen_2012}%
  \BibitemOpen
  \bibfield  {author} {\bibinfo {author} {\bibfnamefont {M.}~\bibnamefont
  {Kortelainen}}, \bibinfo {author} {\bibfnamefont {J.}~\bibnamefont
  {McDonnell}}, \bibinfo {author} {\bibfnamefont {W.}~\bibnamefont
  {Nazarewicz}}, \bibinfo {author} {\bibfnamefont {P.-G.}\ \bibnamefont
  {Reinhard}}, \bibinfo {author} {\bibfnamefont {J.}~\bibnamefont {Sarich}},
  \bibinfo {author} {\bibfnamefont {N.}~\bibnamefont {Schunck}}, \bibinfo
  {author} {\bibfnamefont {M.~V.}\ \bibnamefont {Stoitsov}},\ and\ \bibinfo
  {author} {\bibfnamefont {S.~M.}\ \bibnamefont {Wild}},\ }\href
  {https://doi.org/10.1103/PhysRevC.85.024304} {\bibfield  {journal} {\bibinfo
  {journal} {Phys. Rev. C}\ }\textbf {\bibinfo {volume} {85}},\ \bibinfo
  {pages} {024304} (\bibinfo {year} {2012})}\BibitemShut {NoStop}%
\bibitem [{\citenamefont {Gray}\ \emph {et~al.}(2023)\citenamefont {Gray},
  \citenamefont {Stuchbery}, \citenamefont {Dobaczewski}, \citenamefont
  {Blazhev}, \citenamefont {Alshammari}, \citenamefont {Bignell}, \citenamefont
  {Bonnard}, \citenamefont {Coombes}, \citenamefont {Dowie}, \citenamefont
  {Gerathy}, \citenamefont {Kib\'e{}di}, \citenamefont {Lane}, \citenamefont
  {McCormick}, \citenamefont {Mitchell}, \citenamefont {Nicholls},
  \citenamefont {Pope}, \citenamefont {Reinhard}, \citenamefont {Spinks},\ and\
  \citenamefont {Zhong}}]{Gray_2023}%
  \BibitemOpen
  \bibfield  {author} {\bibinfo {author} {\bibfnamefont {T.~J.}\ \bibnamefont
  {Gray}}, \bibinfo {author} {\bibfnamefont {A.~E.}\ \bibnamefont {Stuchbery}},
  \bibinfo {author} {\bibfnamefont {J.}~\bibnamefont {Dobaczewski}}, \bibinfo
  {author} {\bibfnamefont {A.}~\bibnamefont {Blazhev}}, \bibinfo {author}
  {\bibfnamefont {H.~A.}\ \bibnamefont {Alshammari}}, \bibinfo {author}
  {\bibfnamefont {L.~J.}\ \bibnamefont {Bignell}}, \bibinfo {author}
  {\bibfnamefont {J.}~\bibnamefont {Bonnard}}, \bibinfo {author} {\bibfnamefont
  {B.~J.}\ \bibnamefont {Coombes}}, \bibinfo {author} {\bibfnamefont
  {J.~T.~H.}\ \bibnamefont {Dowie}}, \bibinfo {author} {\bibfnamefont
  {M.~S.~M.}\ \bibnamefont {Gerathy}}, \bibinfo {author} {\bibfnamefont
  {T.}~\bibnamefont {Kib\'e{}di}}, \bibinfo {author} {\bibfnamefont {G.~J.}\
  \bibnamefont {Lane}}, \bibinfo {author} {\bibfnamefont {B.~P.}\ \bibnamefont
  {McCormick}}, \bibinfo {author} {\bibfnamefont {A.~J.}\ \bibnamefont
  {Mitchell}}, \bibinfo {author} {\bibfnamefont {C.}~\bibnamefont {Nicholls}},
  \bibinfo {author} {\bibfnamefont {J.~G.}\ \bibnamefont {Pope}}, \bibinfo
  {author} {\bibfnamefont {P.~G.}\ \bibnamefont {Reinhard}}, \bibinfo {author}
  {\bibfnamefont {N.~J.}\ \bibnamefont {Spinks}},\ and\ \bibinfo {author}
  {\bibfnamefont {Y.}~\bibnamefont {Zhong}},\ }\href@noop {} {\bibinfo {title}
  {Shape polarization in the tin isotopes near $n=60$ from precision $g$-factor
  measurements on short-lived $11/2^-$ isomers}} (\bibinfo {year} {2023}),\
  \Eprint {https://arxiv.org/abs/2310.11980} {arXiv:2310.11980 [nucl-ex]}
  \BibitemShut {NoStop}%
\bibitem [{\citenamefont {Lyutorovich}\ and\ \citenamefont
  {Tselyaev}(2023)}]{Lyutorovich_2023}%
  \BibitemOpen
  \bibfield  {author} {\bibinfo {author} {\bibfnamefont {N.}~\bibnamefont
  {Lyutorovich}}\ and\ \bibinfo {author} {\bibfnamefont {V.}~\bibnamefont
  {Tselyaev}},\ }\href {https://doi.org/10.1142/S0218301323500258} {\bibfield
  {journal} {\bibinfo  {journal} {International Journal of Modern Physics E}\
  }\textbf {\bibinfo {volume} {32}},\ \bibinfo {pages} {2350025} (\bibinfo
  {year} {2023})}\BibitemShut {NoStop}%
\bibitem [{\citenamefont {\'C{}wiok}\ and\ \citenamefont
  {Hofmann}(1994)}]{Cwiok_1994}%
  \BibitemOpen
  \bibfield  {author} {\bibinfo {author} {\bibfnamefont {S.}~\bibnamefont
  {\'C{}wiok}}\ and\ \bibinfo {author} {\bibfnamefont {S.}~\bibnamefont
  {Hofmann}},\ }\href@noop {} {\bibfield  {journal} {\bibinfo  {journal}
  {Nuclear Physics A}\ }\textbf {\bibinfo {volume} {573}},\ \bibinfo {pages}
  {356} (\bibinfo {year} {1994})}\BibitemShut {NoStop}%
\bibitem [{\citenamefont {Parkhomenko}\ and\ \citenamefont
  {Sobiczewski}(2005)}]{Parkhomenko_2005}%
  \BibitemOpen
  \bibfield  {author} {\bibinfo {author} {\bibfnamefont {A.}~\bibnamefont
  {Parkhomenko}}\ and\ \bibinfo {author} {\bibfnamefont {A.}~\bibnamefont
  {Sobiczewski}},\ }\href@noop {} {\bibfield  {journal} {\bibinfo  {journal}
  {Acta Phys. Pol. B}\ }\textbf {\bibinfo {volume} {36}},\ \bibinfo {pages}
  {3115} (\bibinfo {year} {2005})}\BibitemShut {NoStop}%
\bibitem [{\citenamefont {Adamian}\ \emph {et~al.}(2011)\citenamefont
  {Adamian}, \citenamefont {Antonenko}, \citenamefont {Kuklin}, \citenamefont
  {Lu}, \citenamefont {Malov},\ and\ \citenamefont {Zhou}}]{Adamian_2011}%
  \BibitemOpen
  \bibfield  {author} {\bibinfo {author} {\bibfnamefont {G.~G.}\ \bibnamefont
  {Adamian}}, \bibinfo {author} {\bibfnamefont {N.~V.}\ \bibnamefont
  {Antonenko}}, \bibinfo {author} {\bibfnamefont {S.~N.}\ \bibnamefont
  {Kuklin}}, \bibinfo {author} {\bibfnamefont {B.~N.}\ \bibnamefont {Lu}},
  \bibinfo {author} {\bibfnamefont {L.~A.}\ \bibnamefont {Malov}},\ and\
  \bibinfo {author} {\bibfnamefont {S.~G.}\ \bibnamefont {Zhou}},\ }\href
  {https://doi.org/10.1103/PhysRevC.84.024324} {\bibfield  {journal} {\bibinfo
  {journal} {Phys. Rev. C}\ }\textbf {\bibinfo {volume} {84}},\ \bibinfo
  {pages} {024324} (\bibinfo {year} {2011})}\BibitemShut {NoStop}%
\bibitem [{\citenamefont {Shirikova}\ \emph {et~al.}(2015)\citenamefont
  {Shirikova}, \citenamefont {Sushkov}, \citenamefont {Malov},\ and\
  \citenamefont {Jolos}}]{Shirikova_2015}%
  \BibitemOpen
  \bibfield  {author} {\bibinfo {author} {\bibfnamefont {N.~Y.}\ \bibnamefont
  {Shirikova}}, \bibinfo {author} {\bibfnamefont {A.~V.}\ \bibnamefont
  {Sushkov}}, \bibinfo {author} {\bibfnamefont {L.~A.}\ \bibnamefont {Malov}},\
  and\ \bibinfo {author} {\bibfnamefont {R.~V.}\ \bibnamefont {Jolos}},\ }\href
  {https://doi.org/10.1140/epja/i2015-15021-4} {\bibfield  {journal} {\bibinfo
  {journal} {The European Physical Journal A}\ }\textbf {\bibinfo {volume}
  {51}},\ \bibinfo {pages} {21} (\bibinfo {year} {2015})}\BibitemShut {NoStop}%
\bibitem [{\citenamefont {Adamian}\ \emph {et~al.}(2018)\citenamefont
  {Adamian}, \citenamefont {Malov}, \citenamefont {Antonenko},\ and\
  \citenamefont {Jolos}}]{Adamian_2018}%
  \BibitemOpen
  \bibfield  {author} {\bibinfo {author} {\bibfnamefont {G.~G.}\ \bibnamefont
  {Adamian}}, \bibinfo {author} {\bibfnamefont {L.~A.}\ \bibnamefont {Malov}},
  \bibinfo {author} {\bibfnamefont {N.~V.}\ \bibnamefont {Antonenko}},\ and\
  \bibinfo {author} {\bibfnamefont {R.~V.}\ \bibnamefont {Jolos}},\ }\href
  {https://doi.org/10.1103/PhysRevC.97.034308} {\bibfield  {journal} {\bibinfo
  {journal} {Phys. Rev. C}\ }\textbf {\bibinfo {volume} {97}},\ \bibinfo
  {pages} {034308} (\bibinfo {year} {2018})}\BibitemShut {NoStop}%
\bibitem [{\citenamefont {Bender}\ \emph
  {et~al.}(2003{\natexlab{b}})\citenamefont {Bender}, \citenamefont {Bonche},
  \citenamefont {Duguet},\ and\ \citenamefont {Heenen}}]{Bender_NPA_2003}%
  \BibitemOpen
  \bibfield  {author} {\bibinfo {author} {\bibfnamefont {M.}~\bibnamefont
  {Bender}}, \bibinfo {author} {\bibfnamefont {P.}~\bibnamefont {Bonche}},
  \bibinfo {author} {\bibfnamefont {T.}~\bibnamefont {Duguet}},\ and\ \bibinfo
  {author} {\bibfnamefont {P.-H.}\ \bibnamefont {Heenen}},\ }\href
  {https://doi.org/https://doi.org/10.1016/S0375-9474(03)01081-9} {\bibfield
  {journal} {\bibinfo  {journal} {Nuclear Physics A}\ }\textbf {\bibinfo
  {volume} {723}},\ \bibinfo {pages} {354} (\bibinfo {year}
  {2003}{\natexlab{b}})}\BibitemShut {NoStop}%
\bibitem [{\citenamefont {Dobaczewski}\ \emph {et~al.}(2015)\citenamefont
  {Dobaczewski}, \citenamefont {Afanasjev}, \citenamefont {Bender},
  \citenamefont {Robledo},\ and\ \citenamefont {Shi}}]{Dobaczewski_2015}%
  \BibitemOpen
  \bibfield  {author} {\bibinfo {author} {\bibfnamefont {J.}~\bibnamefont
  {Dobaczewski}}, \bibinfo {author} {\bibfnamefont {A.}~\bibnamefont
  {Afanasjev}}, \bibinfo {author} {\bibfnamefont {M.}~\bibnamefont {Bender}},
  \bibinfo {author} {\bibfnamefont {L.}~\bibnamefont {Robledo}},\ and\ \bibinfo
  {author} {\bibfnamefont {Y.}~\bibnamefont {Shi}},\ }\href
  {https://doi.org/https://doi.org/10.1016/j.nuclphysa.2015.07.015} {\bibfield
  {journal} {\bibinfo  {journal} {Nuclear Physics A}\ }\textbf {\bibinfo
  {volume} {944}},\ \bibinfo {pages} {388} (\bibinfo {year} {2015})},\ \bibinfo
  {note} {special Issue on Superheavy Elements}\BibitemShut {NoStop}%
\bibitem [{\citenamefont {Singh}\ \emph {et~al.}(2021)\citenamefont {Singh},
  \citenamefont {Sharma}, \citenamefont {Sharma}, \citenamefont {Kaushik},
  \citenamefont {Jain},\ and\ \citenamefont {Saxena}}]{Singh_2021}%
  \BibitemOpen
  \bibfield  {author} {\bibinfo {author} {\bibfnamefont {U.}~\bibnamefont
  {Singh}}, \bibinfo {author} {\bibfnamefont {R.}~\bibnamefont {Sharma}},
  \bibinfo {author} {\bibfnamefont {P.}~\bibnamefont {Sharma}}, \bibinfo
  {author} {\bibfnamefont {M.}~\bibnamefont {Kaushik}}, \bibinfo {author}
  {\bibfnamefont {S.}~\bibnamefont {Jain}},\ and\ \bibinfo {author}
  {\bibfnamefont {G.}~\bibnamefont {Saxena}},\ }\href
  {https://doi.org/https://doi.org/10.1016/j.nuclphysa.2020.122066} {\bibfield
  {journal} {\bibinfo  {journal} {Nuclear Physics A}\ }\textbf {\bibinfo
  {volume} {1006}},\ \bibinfo {pages} {122066} (\bibinfo {year}
  {2021})}\BibitemShut {NoStop}%
\bibitem [{\citenamefont {Adamian}\ \emph {et~al.}(2021)\citenamefont
  {Adamian}, \citenamefont {Antonenko}, \citenamefont {Lenske},\ and\
  \citenamefont {Malov}}]{Adamian_2021}%
  \BibitemOpen
  \bibfield  {author} {\bibinfo {author} {\bibfnamefont {G.~G.}\ \bibnamefont
  {Adamian}}, \bibinfo {author} {\bibfnamefont {N.~V.}\ \bibnamefont
  {Antonenko}}, \bibinfo {author} {\bibfnamefont {H.}~\bibnamefont {Lenske}},\
  and\ \bibinfo {author} {\bibfnamefont {S.-G.}\ \bibnamefont {Malov},
  \bibfnamefont {L.~A.~Zhou}},\ }\href
  {https://doi.org/10.1140/epja/s10050-021-00375-1} {\bibfield  {journal}
  {\bibinfo  {journal} {The European Physical Journal A}\ }\textbf {\bibinfo
  {volume} {57}},\ \bibinfo {pages} {89} (\bibinfo {year} {2021})}\BibitemShut
  {NoStop}%
\bibitem [{\citenamefont {Shi}\ \emph {et~al.}(2014)\citenamefont {Shi},
  \citenamefont {Dobaczewski},\ and\ \citenamefont {Greenlees}}]{Shi_2014}%
  \BibitemOpen
  \bibfield  {author} {\bibinfo {author} {\bibfnamefont {Y.}~\bibnamefont
  {Shi}}, \bibinfo {author} {\bibfnamefont {J.}~\bibnamefont {Dobaczewski}},\
  and\ \bibinfo {author} {\bibfnamefont {P.~T.}\ \bibnamefont {Greenlees}},\
  }\href {https://doi.org/10.1103/PhysRevC.89.034309} {\bibfield  {journal}
  {\bibinfo  {journal} {Phys. Rev. C}\ }\textbf {\bibinfo {volume} {89}},\
  \bibinfo {pages} {034309} (\bibinfo {year} {2014})}\BibitemShut {NoStop}%
\bibitem [{\citenamefont {Dobaczewski}\ \emph {et~al.}(2021)\citenamefont
  {Dobaczewski}, \citenamefont {B\c{a}czyk}, \citenamefont {Becker},
  \citenamefont {Bender}, \citenamefont {Bennaceur}, \citenamefont {Bonnard},
  \citenamefont {Gao}, \citenamefont {Idini}, \citenamefont {Konieczka},
  \citenamefont {Kortelainen}, \citenamefont {Pr\'ochniak}, \citenamefont
  {Romero}, \citenamefont {Satu\l{}a}, \citenamefont {Shi}, \citenamefont
  {Werner},\ and\ \citenamefont {Yu}}]{Dobaczewski_2021}%
  \BibitemOpen
  \bibfield  {author} {\bibinfo {author} {\bibfnamefont {J.}~\bibnamefont
  {Dobaczewski}}, \bibinfo {author} {\bibfnamefont {P.}~\bibnamefont
  {B\c{a}czyk}}, \bibinfo {author} {\bibfnamefont {P.}~\bibnamefont {Becker}},
  \bibinfo {author} {\bibfnamefont {M.}~\bibnamefont {Bender}}, \bibinfo
  {author} {\bibfnamefont {K.}~\bibnamefont {Bennaceur}}, \bibinfo {author}
  {\bibfnamefont {J.}~\bibnamefont {Bonnard}}, \bibinfo {author} {\bibfnamefont
  {Y.}~\bibnamefont {Gao}}, \bibinfo {author} {\bibfnamefont {A.}~\bibnamefont
  {Idini}}, \bibinfo {author} {\bibfnamefont {M.}~\bibnamefont {Konieczka}},
  \bibinfo {author} {\bibfnamefont {M.}~\bibnamefont {Kortelainen}}, \bibinfo
  {author} {\bibfnamefont {L.}~\bibnamefont {Pr\'ochniak}}, \bibinfo {author}
  {\bibfnamefont {A.~M.}\ \bibnamefont {Romero}}, \bibinfo {author}
  {\bibfnamefont {W.}~\bibnamefont {Satu\l{}a}}, \bibinfo {author}
  {\bibfnamefont {Y.}~\bibnamefont {Shi}}, \bibinfo {author} {\bibfnamefont
  {T.~R.}\ \bibnamefont {Werner}},\ and\ \bibinfo {author} {\bibfnamefont
  {L.~F.}\ \bibnamefont {Yu}},\ }\href
  {https://doi.org/10.1088/1361-6471/ac0a82} {\bibfield  {journal} {\bibinfo
  {journal} {Journal of Physics G: Nuclear and Particle Physics}\ }\textbf
  {\bibinfo {volume} {48}},\ \bibinfo {pages} {102001} (\bibinfo {year}
  {2021})}\BibitemShut {NoStop}%
\bibitem [{\citenamefont {Schunck}(2019)}]{Schunck_2019}%
  \BibitemOpen
  \bibinfo {editor} {\bibfnamefont {N.}~\bibnamefont {Schunck}},\ ed.,\ \href
  {https://doi.org/10.1088/2053-2563/aae0ed} {\emph {\bibinfo {title} {Energy
  Density Functional Methods for Atomic Nuclei}}},\ 2053-2563\ (\bibinfo
  {publisher} {IOP Publishing},\ \bibinfo {year} {2019})\BibitemShut {NoStop}%
\bibitem [{\citenamefont {Ring}\ and\ \citenamefont
  {Schuck}(1980)}]{Ring_Schuck}%
  \BibitemOpen
  \bibfield  {author} {\bibinfo {author} {\bibfnamefont {P.}~\bibnamefont
  {Ring}}\ and\ \bibinfo {author} {\bibfnamefont {P.}~\bibnamefont {Schuck}},\
  }\href@noop {} {\emph {\bibinfo {title} {The Nuclear Many-Body Problem}}}\
  (\bibinfo  {publisher} {Springer--Verl.},\ \bibinfo {address} {New York,
  Heidelberg, Berlin},\ \bibinfo {year} {1980})\BibitemShut {NoStop}%
\bibitem [{nud()}]{nudat3}%
  \BibitemOpen
  \href {https://www.nndc.bnl.gov/nudat3} {\bibinfo  {journal}
  {https://www.nndc.bnl.gov/nudat3}\ }\BibitemShut {NoStop}%
\bibitem [{\citenamefont {Browne}\ and\ \citenamefont
  {Tuli}(2013)}]{A=251-259}%
  \BibitemOpen
\bibfield  {journal} {  }\bibfield  {author} {\bibinfo {author} {\bibfnamefont
  {E.}~\bibnamefont {Browne}}\ and\ \bibinfo {author} {\bibfnamefont
  {J.}~\bibnamefont {Tuli}},\ }\href
  {https://doi.org/https://doi.org/10.1016/j.nds.2013.08.002} {\bibfield
  {journal} {\bibinfo  {journal} {Nuclear Data Sheets}\ }\textbf {\bibinfo
  {volume} {114}},\ \bibinfo {pages} {1041} (\bibinfo {year}
  {2013})}\BibitemShut {NoStop}%
\bibitem [{ENS()}]{ENSDF}%
  \BibitemOpen
  \href {https://www.nndc.bnl.gov/ensdf} {\bibinfo  {journal}
  {https://www.nndc.bnl.gov/ensdf}\ }\BibitemShut {NoStop}%
\bibitem [{\citenamefont {Lopez-Martens}\ \emph {et~al.}(2007)\citenamefont
  {Lopez-Martens}, \citenamefont {Hauschild}, \citenamefont {Yeremin},
  \citenamefont {Dorvaux}, \citenamefont {Belozerov}, \citenamefont
  {Brian\c{c}on}, \citenamefont {Chelnokov}, \citenamefont {Chepigin},
  \citenamefont {Curien}, \citenamefont {D\'esesquelles}, \citenamefont {Gall},
  \citenamefont {Gorshkov}, \citenamefont {Guttormsen}, \citenamefont
  {Hanappe}, \citenamefont {Kabachenko}, \citenamefont {Khalfallah},
  \citenamefont {Korichi}, \citenamefont {Larsen}, \citenamefont {Malyshev},
  \citenamefont {Minkova}, \citenamefont {Oganessian}, \citenamefont {Popeko},
  \citenamefont {Rousseau}, \citenamefont {Rowley}, \citenamefont {Sagaidak},
  \citenamefont {Sharo}, \citenamefont {Shutov}, \citenamefont {Siem},
  \citenamefont {Stuttg\'e}, \citenamefont {Svirikhin}, \citenamefont {Syed},
  ,\ and\ \citenamefont {Theisen}}]{Lopez-Martens_2007}%
  \BibitemOpen
\bibfield  {journal} {  }\bibfield  {author} {\bibinfo {author} {\bibfnamefont
  {A.}~\bibnamefont {Lopez-Martens}}, \bibinfo {author} {\bibfnamefont
  {K.}~\bibnamefont {Hauschild}}, \bibinfo {author} {\bibfnamefont
  {A.}~\bibnamefont {Yeremin}}, \bibinfo {author} {\bibfnamefont
  {O.}~\bibnamefont {Dorvaux}}, \bibinfo {author} {\bibfnamefont
  {A.}~\bibnamefont {Belozerov}}, \bibinfo {author} {\bibfnamefont
  {C.}~\bibnamefont {Brian\c{c}on}}, \bibinfo {author} {\bibfnamefont
  {M.}~\bibnamefont {Chelnokov}}, \bibinfo {author} {\bibfnamefont
  {V.}~\bibnamefont {Chepigin}}, \bibinfo {author} {\bibfnamefont
  {D.}~\bibnamefont {Curien}}, \bibinfo {author} {\bibfnamefont
  {P.}~\bibnamefont {D\'esesquelles}}, \bibinfo {author} {\bibfnamefont
  {B.}~\bibnamefont {Gall}}, \bibinfo {author} {\bibfnamefont {V.}~\bibnamefont
  {Gorshkov}}, \bibinfo {author} {\bibfnamefont {M.}~\bibnamefont
  {Guttormsen}}, \bibinfo {author} {\bibfnamefont {F.}~\bibnamefont {Hanappe}},
  \bibinfo {author} {\bibfnamefont {A.}~\bibnamefont {Kabachenko}}, \bibinfo
  {author} {\bibfnamefont {F.}~\bibnamefont {Khalfallah}}, \bibinfo {author}
  {\bibfnamefont {A.}~\bibnamefont {Korichi}}, \bibinfo {author} {\bibfnamefont
  {A.}~\bibnamefont {Larsen}}, \bibinfo {author} {\bibfnamefont
  {O.}~\bibnamefont {Malyshev}}, \bibinfo {author} {\bibfnamefont
  {A.}~\bibnamefont {Minkova}}, \bibinfo {author} {\bibfnamefont
  {Y.}~\bibnamefont {Oganessian}}, \bibinfo {author} {\bibfnamefont
  {A.}~\bibnamefont {Popeko}}, \bibinfo {author} {\bibfnamefont
  {M.}~\bibnamefont {Rousseau}}, \bibinfo {author} {\bibfnamefont
  {N.}~\bibnamefont {Rowley}}, \bibinfo {author} {\bibfnamefont
  {R.}~\bibnamefont {Sagaidak}}, \bibinfo {author} {\bibfnamefont
  {S.}~\bibnamefont {Sharo}}, \bibinfo {author} {\bibfnamefont
  {A.}~\bibnamefont {Shutov}}, \bibinfo {author} {\bibfnamefont
  {S.}~\bibnamefont {Siem}}, \bibinfo {author} {\bibfnamefont {L.}~\bibnamefont
  {Stuttg\'e}}, \bibinfo {author} {\bibfnamefont {A.}~\bibnamefont
  {Svirikhin}}, \bibinfo {author} {\bibfnamefont {N.}~\bibnamefont {Syed}}, ,\
  and\ \bibinfo {author} {\bibfnamefont {C.}~\bibnamefont {Theisen}},\ }\href
  {https://doi.org/DOI 10.1140/epja/i2007-10391-8} {\bibfield  {journal}
  {\bibinfo  {journal} {Eur. Phys. J. A}\ }\textbf {\bibinfo {volume} {32}},\
  \bibinfo {pages} {245} (\bibinfo {year} {2007})}\BibitemShut {NoStop}%
\bibitem [{\citenamefont {Hauschild}\ \emph {et~al.}(2022)\citenamefont
  {Hauschild}, \citenamefont {Lopez-Martens}, \citenamefont {Chakma},
  \citenamefont {Chelnokov}, \citenamefont {Chepigin}, \citenamefont {Isaev},
  \citenamefont {Izosimov}, \citenamefont {Katrasev}, \citenamefont
  {Kuznetsova}, \citenamefont {Malyshev}, \citenamefont {Popeko}, \citenamefont
  {Popov}, \citenamefont {Sokol}, \citenamefont {Svirikhin}, \citenamefont
  {Tezekbayeva}, \citenamefont {Yeremin}, \citenamefont {Asfari}, \citenamefont
  {Dorvaux}, \citenamefont {Gall}, \citenamefont {Kessaci}, \citenamefont
  {Ackermann}, \citenamefont {Piot}, \citenamefont {Mosat},\ and\ \citenamefont
  {Andel}}]{Hauschild_2022}%
  \BibitemOpen
  \bibfield  {author} {\bibinfo {author} {\bibfnamefont {K.}~\bibnamefont
  {Hauschild}}, \bibinfo {author} {\bibfnamefont {A.}~\bibnamefont
  {Lopez-Martens}}, \bibinfo {author} {\bibfnamefont {R.}~\bibnamefont
  {Chakma}}, \bibinfo {author} {\bibfnamefont {M.~L.}\ \bibnamefont
  {Chelnokov}}, \bibinfo {author} {\bibfnamefont {V.~I.}\ \bibnamefont
  {Chepigin}}, \bibinfo {author} {\bibfnamefont {A.~V.}\ \bibnamefont {Isaev}},
  \bibinfo {author} {\bibfnamefont {I.~N.}\ \bibnamefont {Izosimov}}, \bibinfo
  {author} {\bibfnamefont {D.~E.}\ \bibnamefont {Katrasev}}, \bibinfo {author}
  {\bibfnamefont {A.~A.}\ \bibnamefont {Kuznetsova}}, \bibinfo {author}
  {\bibfnamefont {O.~N.}\ \bibnamefont {Malyshev}}, \bibinfo {author}
  {\bibfnamefont {A.~G.}\ \bibnamefont {Popeko}}, \bibinfo {author}
  {\bibfnamefont {Y.~A.}\ \bibnamefont {Popov}}, \bibinfo {author}
  {\bibfnamefont {E.~A.}\ \bibnamefont {Sokol}}, \bibinfo {author}
  {\bibfnamefont {A.~I.}\ \bibnamefont {Svirikhin}}, \bibinfo {author}
  {\bibfnamefont {M.~S.}\ \bibnamefont {Tezekbayeva}}, \bibinfo {author}
  {\bibfnamefont {A.~V.}\ \bibnamefont {Yeremin}}, \bibinfo {author}
  {\bibfnamefont {Z.}~\bibnamefont {Asfari}}, \bibinfo {author} {\bibfnamefont
  {O.}~\bibnamefont {Dorvaux}}, \bibinfo {author} {\bibfnamefont {B.~J.~P.}\
  \bibnamefont {Gall}}, \bibinfo {author} {\bibfnamefont {K.}~\bibnamefont
  {Kessaci}}, \bibinfo {author} {\bibfnamefont {D.}~\bibnamefont {Ackermann}},
  \bibinfo {author} {\bibfnamefont {J.}~\bibnamefont {Piot}}, \bibinfo {author}
  {\bibfnamefont {P.}~\bibnamefont {Mosat}},\ and\ \bibinfo {author}
  {\bibfnamefont {B.}~\bibnamefont {Andel}},\ }\bibfield  {journal} {\bibinfo
  {journal} {The European Physical Journal A}\ }\textbf {\bibinfo {volume}
  {58}},\ \href {https://doi.org/10.1140/epja/s10050-021-00657-8}
  {10.1140/epja/s10050-021-00657-8} (\bibinfo {year} {2022})\BibitemShut
  {NoStop}%
\bibitem [{\citenamefont {F.P.He\ss{}berger}\ \emph {et~al.}(1997)\citenamefont
  {F.P.He\ss{}berger}, \citenamefont {S.Hofmann}, \citenamefont {V.Ninov},
  \citenamefont {P.Armbruster}, \citenamefont {H.Folger}, \citenamefont
  {G.M\"unzenberg}, \citenamefont {H.J.Sch\"ott}, \citenamefont {A.G.Popeko},
  \citenamefont {A.V.Yeremin}, \citenamefont {A.N.Andreyev},\ and\
  \citenamefont {S.Saro}}]{Hessberger_1997}%
  \BibitemOpen
  \bibfield  {author} {\bibinfo {author} {\bibnamefont {F.P.He\ss{}berger}},
  \bibinfo {author} {\bibnamefont {S.Hofmann}}, \bibinfo {author} {\bibnamefont
  {V.Ninov}}, \bibinfo {author} {\bibnamefont {P.Armbruster}}, \bibinfo
  {author} {\bibnamefont {H.Folger}}, \bibinfo {author} {\bibnamefont
  {G.M\"unzenberg}}, \bibinfo {author} {\bibnamefont {H.J.Sch\"ott}}, \bibinfo
  {author} {\bibnamefont {A.G.Popeko}}, \bibinfo {author} {\bibnamefont
  {A.V.Yeremin}}, \bibinfo {author} {\bibnamefont {A.N.Andreyev}},\ and\
  \bibinfo {author} {\bibnamefont {S.Saro}},\ }\href@noop {} {\bibfield
  {journal} {\bibinfo  {journal} {Z. Phys. A}\ }\textbf {\bibinfo {volume}
  {359}},\ \bibinfo {pages} {415} (\bibinfo {year} {1997})}\BibitemShut
  {NoStop}%
\bibitem [{\citenamefont {Morse}(2023)}]{A=251}%
  \BibitemOpen
  \bibfield  {author} {\bibinfo {author} {\bibfnamefont {C.}~\bibnamefont
  {Morse}},\ }\href {https://doi.org/https://doi.org/10.1016/j.nds.2023.04.002}
  {\bibfield  {journal} {\bibinfo  {journal} {Nuclear Data Sheets}\ }\textbf
  {\bibinfo {volume} {189}},\ \bibinfo {pages} {111} (\bibinfo {year}
  {2023})}\BibitemShut {NoStop}%
\bibitem [{\citenamefont {Lopez-Martens}\ \emph {et~al.}(2022)\citenamefont
  {Lopez-Martens}, \citenamefont {Hauschild}, \citenamefont {Svirikhin},
  \citenamefont {Asfari}, \citenamefont {Chelnokov}, \citenamefont {Chepigin},
  \citenamefont {Dorvaux}, \citenamefont {Forge}, \citenamefont {Gall},
  \citenamefont {Isaev}, \citenamefont {Izosimov}, \citenamefont {Kessaci},
  \citenamefont {Kuznetsova}, \citenamefont {Malyshev}, \citenamefont {Mukhin},
  \citenamefont {Popeko}, \citenamefont {Popov}, \citenamefont {Sailaubekov},
  \citenamefont {Sokol}, \citenamefont {Tezekbayeva},\ and\ \citenamefont
  {Yeremin}}]{Lopez-Martens_2022}%
  \BibitemOpen
  \bibfield  {author} {\bibinfo {author} {\bibfnamefont {A.}~\bibnamefont
  {Lopez-Martens}}, \bibinfo {author} {\bibfnamefont {K.}~\bibnamefont
  {Hauschild}}, \bibinfo {author} {\bibfnamefont {A.~I.}\ \bibnamefont
  {Svirikhin}}, \bibinfo {author} {\bibfnamefont {Z.}~\bibnamefont {Asfari}},
  \bibinfo {author} {\bibfnamefont {M.~L.}\ \bibnamefont {Chelnokov}}, \bibinfo
  {author} {\bibfnamefont {V.~I.}\ \bibnamefont {Chepigin}}, \bibinfo {author}
  {\bibfnamefont {O.}~\bibnamefont {Dorvaux}}, \bibinfo {author} {\bibfnamefont
  {M.}~\bibnamefont {Forge}}, \bibinfo {author} {\bibfnamefont
  {B.}~\bibnamefont {Gall}}, \bibinfo {author} {\bibfnamefont {A.~V.}\
  \bibnamefont {Isaev}}, \bibinfo {author} {\bibfnamefont {I.~N.}\ \bibnamefont
  {Izosimov}}, \bibinfo {author} {\bibfnamefont {K.}~\bibnamefont {Kessaci}},
  \bibinfo {author} {\bibfnamefont {A.~A.}\ \bibnamefont {Kuznetsova}},
  \bibinfo {author} {\bibfnamefont {O.~N.}\ \bibnamefont {Malyshev}}, \bibinfo
  {author} {\bibfnamefont {R.~S.}\ \bibnamefont {Mukhin}}, \bibinfo {author}
  {\bibfnamefont {A.~G.}\ \bibnamefont {Popeko}}, \bibinfo {author}
  {\bibfnamefont {Y.~A.}\ \bibnamefont {Popov}}, \bibinfo {author}
  {\bibfnamefont {B.}~\bibnamefont {Sailaubekov}}, \bibinfo {author}
  {\bibfnamefont {E.~A.}\ \bibnamefont {Sokol}}, \bibinfo {author}
  {\bibfnamefont {M.~S.}\ \bibnamefont {Tezekbayeva}},\ and\ \bibinfo {author}
  {\bibfnamefont {A.~V.}\ \bibnamefont {Yeremin}},\ }\href
  {https://doi.org/10.1103/PhysRevC.105.L021306} {\bibfield  {journal}
  {\bibinfo  {journal} {Phys. Rev. C}\ }\textbf {\bibinfo {volume} {105}},\
  \bibinfo {pages} {L021306} (\bibinfo {year} {2022})}\BibitemShut {NoStop}%
\bibitem [{\citenamefont {Tezekbayeva}\ \emph {et~al.}(2022)\citenamefont
  {Tezekbayeva}, \citenamefont {Yeremin}, \citenamefont {Svirikhin},
  \citenamefont {Lopez-Martens}, \citenamefont {Chelnokov}, \citenamefont
  {Chepigin}, \citenamefont {Isaev}, \citenamefont {Izosimov}, \citenamefont
  {Karpov}, \citenamefont {Kuznetsova}, \citenamefont {Malyshev}, \citenamefont
  {Mukhin}, \citenamefont {Popeko}, \citenamefont {Popov}, \citenamefont
  {Rachkov}, \citenamefont {Sailaubekov}, \citenamefont {Sokol}, \citenamefont
  {Hauschild}, \citenamefont {Jacob}, \citenamefont {Chakma}, \citenamefont
  {Dorvaux}, \citenamefont {Forge}, \citenamefont {Gall}, \citenamefont
  {Kessaci}, \citenamefont {Andel}, \citenamefont {Antalic}, \citenamefont
  {Bronis},\ and\ \citenamefont {Mosat}}]{Tezekbayeva_2022}%
  \BibitemOpen
  \bibfield  {author} {\bibinfo {author} {\bibfnamefont {M.~S.}\ \bibnamefont
  {Tezekbayeva}}, \bibinfo {author} {\bibfnamefont {A.~V.}\ \bibnamefont
  {Yeremin}}, \bibinfo {author} {\bibfnamefont {A.~I.}\ \bibnamefont
  {Svirikhin}}, \bibinfo {author} {\bibfnamefont {A.}~\bibnamefont
  {Lopez-Martens}}, \bibinfo {author} {\bibfnamefont {M.~L.}\ \bibnamefont
  {Chelnokov}}, \bibinfo {author} {\bibfnamefont {V.~I.}\ \bibnamefont
  {Chepigin}}, \bibinfo {author} {\bibfnamefont {A.~V.}\ \bibnamefont {Isaev}},
  \bibinfo {author} {\bibfnamefont {I.~N.}\ \bibnamefont {Izosimov}}, \bibinfo
  {author} {\bibfnamefont {A.~V.}\ \bibnamefont {Karpov}}, \bibinfo {author}
  {\bibfnamefont {A.~A.}\ \bibnamefont {Kuznetsova}}, \bibinfo {author}
  {\bibfnamefont {O.~N.}\ \bibnamefont {Malyshev}}, \bibinfo {author}
  {\bibfnamefont {R.~S.}\ \bibnamefont {Mukhin}}, \bibinfo {author}
  {\bibfnamefont {A.~G.}\ \bibnamefont {Popeko}}, \bibinfo {author}
  {\bibfnamefont {Y.~A.}\ \bibnamefont {Popov}}, \bibinfo {author}
  {\bibfnamefont {V.~A.}\ \bibnamefont {Rachkov}}, \bibinfo {author}
  {\bibfnamefont {B.~S.}\ \bibnamefont {Sailaubekov}}, \bibinfo {author}
  {\bibfnamefont {E.~A.}\ \bibnamefont {Sokol}}, \bibinfo {author}
  {\bibfnamefont {K.}~\bibnamefont {Hauschild}}, \bibinfo {author}
  {\bibfnamefont {H.}~\bibnamefont {Jacob}}, \bibinfo {author} {\bibfnamefont
  {R.}~\bibnamefont {Chakma}}, \bibinfo {author} {\bibfnamefont
  {O.}~\bibnamefont {Dorvaux}}, \bibinfo {author} {\bibfnamefont
  {M.}~\bibnamefont {Forge}}, \bibinfo {author} {\bibfnamefont
  {B.}~\bibnamefont {Gall}}, \bibinfo {author} {\bibfnamefont {K.}~\bibnamefont
  {Kessaci}}, \bibinfo {author} {\bibfnamefont {B.}~\bibnamefont {Andel}},
  \bibinfo {author} {\bibfnamefont {S.}~\bibnamefont {Antalic}}, \bibinfo
  {author} {\bibfnamefont {A.}~\bibnamefont {Bronis}},\ and\ \bibinfo {author}
  {\bibfnamefont {P.}~\bibnamefont {Mosat}},\ }\bibfield  {journal} {\bibinfo
  {journal} {The European Physical Journal A}\ }\textbf {\bibinfo {volume}
  {58}},\ \href {https://doi.org/10.1140/epja/s10050-022-00707-9}
  {10.1140/epja/s10050-022-00707-9} (\bibinfo {year} {2022})\BibitemShut
  {NoStop}%
\end{thebibliography}%


%
\end{document}